# Facile synthesis of 2D graphene oxide sheet enveloping ultrafine 1D $LiMn_2O_4$ as interconnected framework to enhance cathodic property for Li-ion battery

Niraj Kumar[a], Jassiel R. Rodriguez[b], Vilas G. Pol[b*], Arijit Sen[a*]

[a]SRM Research Institute and Department of Physics & Nanotechnology, SRM Institute of Science & Technology, Kattankulathur 603203, India.

[b]Davidson School of Chemical Engineering, Purdue University, West Lafayette, Indiana 47907, United States.

*corresponding author: vpol@purdue.edu; arijit.s@res.srmuniv.ac.in

## Abstract

Cubic spinel lithium manganese oxide ($LiMn_2O_4$) has been able to attract a great deal of attention over the years as a promising cathode material for large-scale lithium-ion batteries. Here a facile hydrothermal route followed by solid state reaction is developed using as grown ultrafine α-$MnO_2$ nanorods to prepare 1D $LiMn_2O_4$ with 10-50 nm diameters. To enhance the cathodic property of these nanorods, a unique synthesis technique of heat treatment is developed to grow 2D graphene oxide sheet enveloping 1D $LiMn_2O_4$ as interconnected framework. This nanocomposite 3D porous cathode exhibits a high specific charge capacity of 130 mAh $g^{-1}$ at 0.05 C rate and Coulombic efficiency of ~98% after 100 cycles in the potential window of 3.5 to 4.3 V vs Li/$Li^+$ with promising initial charge capacity retention of ~87%, and outstanding structural stability even after 100 cycles. Enhancement in the lithiation and de-lithiation processes leading to improved performance is likely to have its origin in the 2D conducting graphene oxide sheets. It allows for decreasing the Mn dissolution, improve the electron conductivity and reduce the Li-ion path diffusion inside the favourable morphology and crystallinity of the ultrafine 1D $LiMn_2O_4$ nanorods, giving rise to a promising cathode nanocomposite.

## 1. Introduction

Li-ion batteries (LIBs) in recent time are used extensively as the main source of power in almost every portable electronic device like personal computers, phones, digital cameras and electric vehicles, which is due to their compact size and high portability [1].

LIBs have naturally emerged as one of the most promising sources of energy storage, which can potentially replace the non-renewable energy resources like gasoline [2, 3]. However, to meet this end, high energy and power are expected from LIBs, and therefore, cathode materials with high Li-ion storage capacity are of great interest [4-6]. $LiCoO_2$ [7], $LiNi_{1-x-y}Co_xMn_yO_2$ [8], $LiFePO_4$ [9] and $LiMn_2O_4$ [2, 3] have been tested as some of the most promising cathode materials. Among these, $LiMn_2O_4$ is favoured because this offers fast kinetics of Li-ions into its interstitial sites due to its open three-dimensional (3D) framework structure [10-11]. Moreover, it requires low cost precursors such as Manganese (Mn), which is the twelveth most abundant element in earth's crust [11]. It also offers environmental friendliness, high working potential and a promising energy density [12-16]. It is assumed that nanoscale materials of $LiMn_2O_4$ could achieve better cathode performance than its bulk counterpart because these offer shorter $Li^+$-ion path diffusion length, which is due to its particle size decrease but surface area increase, favouring an increment of the electrolyte/electrode interface area [17-19]. In this regard, diverse nanoscale morphologies have been reported, such as nanotubes [20], nanorods [21], nanocones [22], nanochains [14] and nanospheres [23]. Normally, a high quality one-dimensional (1D) nanostructure (nanorod or nanowire) is the desired choice, as its minimal diameters provides an efficient Li-ion conductive pathway without much sacrifice to the volume change [24-28]. Xie *et al*. [11] and Kebede *et al*. [29] have reported superior cyclability for $LiMn_2O_4$ nanorods as LIBs cathode material.

Unfortunately, $LiMn_2O_4$ (LMO) possesses some serious drawbacks that plague its commercial application as Li-ion battery cathode. Poor electronic and ionic conductivities, severe manganese dissolution inside the electrolyte [30-33], and surface reactions caused by Jahn-Teller distortions [34-36] remain some of the principal issues that lead to continuous capacity fading, poor cyclabilty as well as stability of LMO, which need to be addressed to develop a promising cathode material. Carbon coating over LMO can be an alternative for enhancing the performance of LMO electrodes [37, 12-14]. Mesoporous carbon materials have good electronic and ionic conductivities and are thus effective in reducing the $Li^+$-diffusion pathway, which can enhance the rate capability [38-42]. In addition, carbon coating can safeguard the material from unwanted surface distortions and physical change that is highly beneficial for improving the cyclic stability [43, 44]. Graphene is preferred as carbonaceous material for its high electronic and ionic conductivities and its high surface area as a result of its two-dimensional (2D) structure [45-50]. He *et al*. [51] and Ho *et al*. [52]

have recently reported the use of graphene with LMO for developing high performance cathode materials.

In the current work, we have developed ultrafine 1D $LiMn_2O_4$ nanorods of about 10-50 nm diameter using $α-MnO_2$ nanorod as partial template. Initially, ultrafine 1D $α-MnO_2$ nanorods were synthesized using a hydrothermal methodology, as reported previously [53]. Then, 1D $α-MnO_2$ nanorods were mixed with LiOH through hydrothermal technique. Finally, ultrafine $LiMn_2O_4$ nanorods were developed by giving a heat treatment to the mixture. This is one of the few reports that illustrate the hydrothermal technique [53-56] for an efficient and homogeneous mixing of precursors, which lead to develop successfully the ultrafine 1D $LiMn_2O_4$ nanorods. Furthermore, we have developed a simple and unique route to wrap the ultrafine 1D $LiMn_2O_4$ nanorods with graphene oxide by simply mixing under continuous stirring, followed by a heat treatment at 750°C. A promising morphology comprising an interconnected framework of graphene oxide nanosheet connecting the nanorods is presented. The electrochemical evaluation of the synthesized materials as cathode in Li-ion battery reveals that the as prepared $LiMn_2O_4$/graphene oxide nanocomposite delivers a superior cathode performance in comparison with the pristine ultrafine 1D $LiMn_2O_4$.

## 2. Experimental

### 2.1 Reagents

Highly pure Potassium permanganate ($KMnO_4$), Sodium nitrite ($NaNO_2$), Sulphuric acid ($H_2SO_4$) and Lithium hydroxide monohydrate ($LiOH•H_2O$) were received from Sigma Aldrich. Graphite having purity of 96.99% was purchased from United Nanotech Innovation Private Limited. All the solutions were prepared with de-ionized water.

### 2.2 Synthesis of $LiMn_2O_4$ nanorods

36 mg of $LiOH•H_2O$ and 146 mg of as prepared α-$MnO_2$ nanorods were mixed in 40 ml of de-ionized water. α-$MnO_2$ nanorods were prepared following our previous reports [55, 56] with some modifications as presented in supplementary material. The solution was magnetically stirred for 1 h to make a homogenous solution. Then, the solution was transferred to a Teflon-lined autoclave of 50 ml, which was kept in a muffle furnace at 170°C for 12 h. The final solution was dried under continuous stirring at 80°C. The product obtained was grinded in agate mortar and was kept in alumina crucible for solid-state reaction at 750°C for 24 h to obtain ultrafine 1D $LiMn_2O_4$ nanorods.

## 2.3 Synthesis of LiMn$_2$O$_4$/graphene oxide nanocomposite

In the supplementary information, we present the details of graphene oxide sheets preparation through a revised Hummer's method [57]. 2 mg of graphene oxide sheets were grinded with 40 mg of the as prepared ultrafine 1D LiMn$_2$O$_4$ nanorods. Then, 40 ml of deionized water were added into the mixture, which was stirred for 1h to make a homogeneous solution. Following this, the resultant solution was dried by slow heating at 80°C. Next, the as obtained product was grinded, and finally calcinated at 750°C for 24 h to get LiMn$_2$O$_4$/graphene oxide (LMO/GO) nanocomposite.

## 2.4 Characterization

The structures of the as prepared samples were evaluated using X-ray diffraction (XRD) technique through a PAN analytical X' Pert Pro diffractometer. Further structural analysis was followed by Raman spectroscopy using LASER Raman spectrometer. The elemental composition of the samples was determined by X-ray photoelectron (XPS) spectroscopy using 'K-Alpha Instruments, USA' with 400 μm sized aluminium as metal source. The topographies were observed through field emission scanning electron microscope (Quanta 200 FEG FE-SEM) and high-resolution transmission electron microscope (FEI-Tecnai G2 20 S-TWIN).

## 2.5 Electrochemical measurements

Working Li-ion battery electrodes were prepared with 80 wt% active material, 10 wt% Super P, and 10 wt% polyvinylidene fluoride (PVDF, Kynar 310F) on an Al foil. First these materials were dispersed in deionized water and then milled for 30 min. The resulting slurry was deposited on an Al foil, resulting in a thin film, where the loading density of the active material was about 2 mg cm$^{-2}$. After, drying under vacuum at 80 °C for 12 h, the Al foil was cut into circular discs, which were used to fabricate coin-type 2032 cells in an Ar-filled glovebox. Also, for the fabrication of the cell, it was used Li foil as counter electrode, discs of Celgard™ 2500 as separator, and a solution of 1.0 M LiPF$_6$ in Ethylene carbonate (EC) and Diethyl carbonate (DEC) (1:1 volume ratio) as electrolyte. The cells were tested by cyclic voltammetry and galvanostatic charge-discharge cycling. The CV were obtained from an Gamry 600 potentiostat/galvanostat/ZRA at a scan rate of 0.1 mV s$^{-1}$, within a voltage window of 3.0–4.3 V vs Li/Li$^+$. The galvanostatic charge-discharge cycles were conducted between 3.0 to 4.3 V vs Li/Li$^+$ on a battery cycler (*Espec*) at 25 °C.

# 3 Results and discussions

## 3.1 Structural analysis

Figure 1a shows the XRD patterns of the samples obtained using $\alpha$-MnO$_2$ nanorods and LiOH•H$_2$O as precursors. The diffraction peaks observed at 2θ position of 18.688, 36.077, 37.935, 44.01, 48.233, 58.194, 63.9 and 67.311, correspond to Miller indices of (111), (311), (222), (400), (331), (511), (440) and (531), respectively, which agree well with JCPDS card number 88-1026, revealing a cubic spinel phase for the ultrafine 1D LiMn$_2$O$_4$ nanorods. No peaks for any impurity are visible in the spectrum, which suggests that the as prepared LMO NRs are highly pure. In Fig. 1a, the XRD pattern of the LiMn$_2$O$_4$/GO nanocomposite exhibits an additional peak at 2θ position of 26.6 that corresponds to miller indices of (002) related to the graphene oxide sheets, which is in tune with JCPDS card number 89-7213.

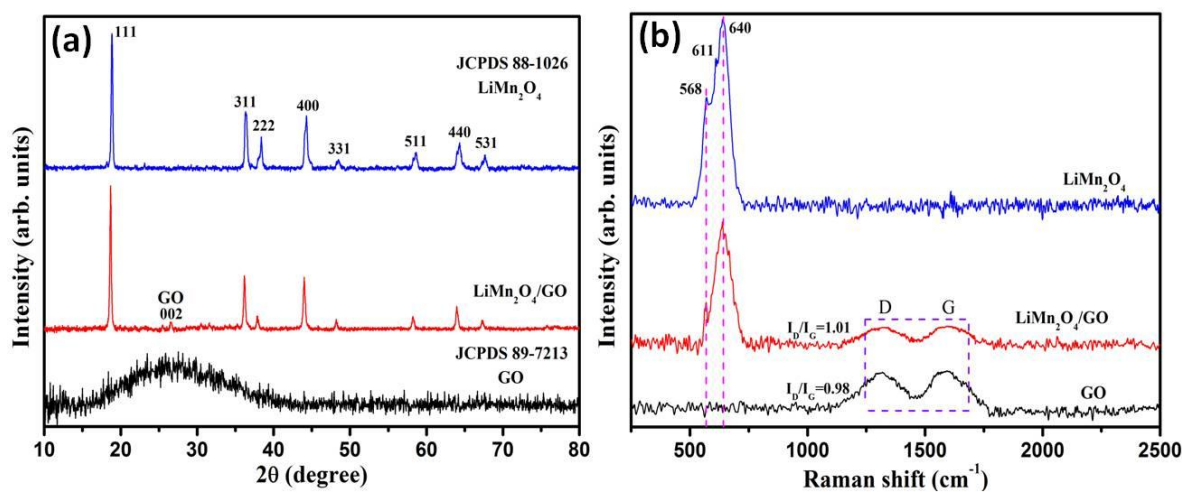

**Figure 1** (a) XRD patterns and (b) Raman spectra of the ultrafine 1D LiMn$_2$O$_4$ nanorods, LiMn$_2$O$_4$/GO nanocomposite and GO.

Fig. 1b portrays the Raman spectra of ultrafine 1D LiMn$_2$O$_4$ nanorods, LiMn$_2$O$_4$/GO nanocomposite and GO sheets. Where, a strong active band can be observed near 640 cm$^{-1}$ along with some weaker active bands at around 611 and 568 cm$^{-1}$. The bands at around 640 and 611 cm$^{-1}$ corresponds to active vibrations of oxygen atoms in the spinel structure oxide in the octahedral MnO$_6$ unit [58, 59]. These bands can be further assigned to Mn-O stretching vibration inside MnO$_6$ [58] relating to $A_{1g}$ species with the $O_h^7$ spectroscopic symmetry. The band at around 568 cm$^{-1}$ can be seen as shoulder peak for its low intensity, which is in close proximity to lithium stoichiometry and so it can be related to Mn$^{IV}$–O vibrations [58]. The weak peak observed at 611 cm$^{-1}$ is not seen in the spectrum of LiMn$_2$O$_4$/GO nanocomposite (Fig. 1b) accounting for an interaction between the ultrafine LiMn$_2$O$_4$ nanorods and graphene oxide sheets. GO is characterized by the presence of D and G band at 1593 and 1320 cm$^{-1}$, which can be seen in Fig. 1b [60]. The LiMn$_2$O$_4$/GO nanocomposite also exhibits the D and G band likewise attributing to the presence of graphene oxide [60]. The intensity of D band (I$_D$) with respect to G band (I$_G$) in the composite was higher as against the graphene oxide sample as I$_D$/I$_G$ ratio was calculated to be 1.01 and 0.98, respectively. This reveals higher amount of disorder in sp$^2$ hybridized carbon atoms of the composite due to their conversion to sp$^3$ hybridization as an effect from the oxygen atoms of LiMn$_2$O$_4$ [61, 62]. From, our results, a favourable bond between the as prepared ultrafine 1D LiMn$_2$O$_4$ nanorods and graphene oxide sheets can be predicted.

### 3.2 Elemental analysis

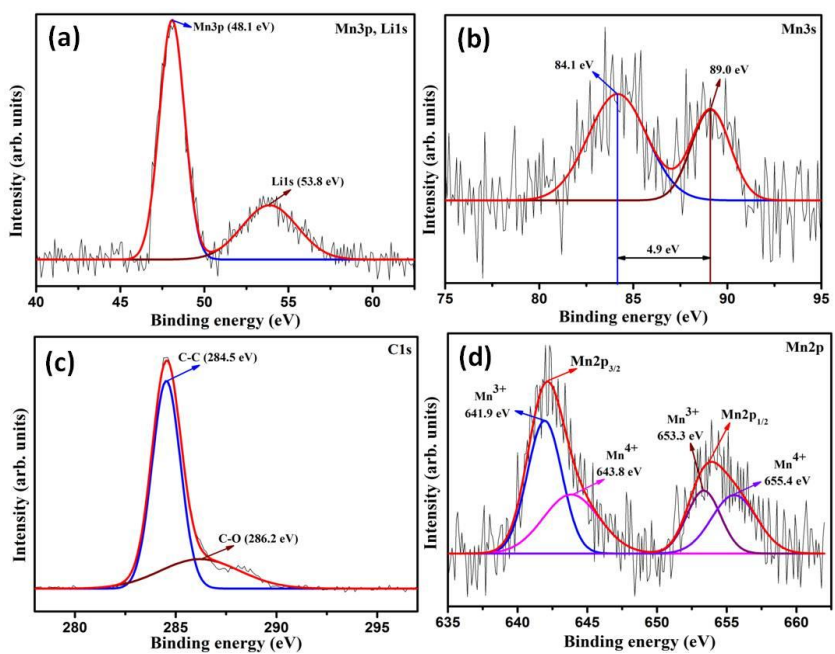

**Figure 2** (a) Mn3p, Li1s, (b) Mn3s, (c) C1s and (d) Mn2p XPS spectra of LiMn$_2$O$_4$/GO nanocomposite.

XPS analysis was carried out to confirm the presence of different elements and their oxidation states in the as prepared nanocomposite. Figure S4a (Supplementary material) reports a wide scan from 0-800 eV, revealing the presence of Li, Mn, C and O elements through the observed photoemission peaks at binding energies of 48.1, 53.8, 84.1, 284.5, 529.7, 641.9 and 653.3 eV for Mn3$p$, Li1$s$, Mn3$s$, C1$s$, O1$s$, Mn2$p_{3/2}$ and Mn2$p_{1/2}$, respectively [62-66]. The as prepared sample shows high purity as no peaks for other impurity elements are present. The high intense peak of C1s is attributed to the presence of carbon from graphene oxide [67]. The concentrated photoemission spectrum for Mn3p and Li1s from 40-60 eV (Fig. 2a) clearly shows two peaks. The stronger one at 48.1 eV corresponds to the Mn3p core level whilst the weaker one at 53.8 eV corresponds to Li1s. Pristine Li metal exhibits a little higher Li1s core level photoemission peak (54.8 eV), which suggests for the existence of Li element as $Li^+$-ions in the as prepared sample [58]. Figure 2b portrays the photoemission spectrum of Mn3s core level. Two split-up peaks can be seen at 84.1 and 89.0 eV with a binding energy difference, $\Delta E$ = 4.9 eV. At the time of photoelectron ejection, the 3s electrons are parallel spin coupled with 3d electrons causing this peak splitting of Mn3s from its satellite, which is useful for evaluating the oxidation state of Mn [63]. The oxidation state of Mn element was calculated to be +3.44 from the following mathematical interpretation: $OS$ = 8.9561 − 1.126 ($\Delta E$), where $OS$ represents average oxidation state [68]. The C1s spectrum displayed in Fig. 2c, shows one prominent peak at 284.5 eV and another less intense peak at 286.2 eV, corresponding to $sp^3$, $sp^2$ hybridized C–C and C–O bonds from graphene oxide [67]. A prominent photoemission peak detected at 529.7 eV in O1s spectrum (Fig. S4b, supplementary information) is ascribed to the $O^{2-}$-ions [64]. The small peak at 532.7 eV can be attributed to the presence of adsorbed impurity from the surrounding moisture content ($OH^-$) [64]. Spin orbit splitting can be seen in Mn2p spectrum (Fig. 2d), attributing to $Mn2p_{3/2}$ and $Mn2p_{1/2}$ core levels at 642.2 and 653.8 eV, respectively, with an energy separation of 11.6 eV [69, 70]. The $Mn2p_{3/2}$ core level peak is deconvoluted to two peaks at 641.9 and 643.8 eV, highlighting the respective +3 (bixbyite) and +4 (pyrolusite) oxidation states of Mn in $LiMn_2^{3+,4+}O_4$ [69, 70]. Likewise, the two deconvoluted peaks of $Mn2p_{1/2}$ core level at 653.3 and 655.4 eV correspond to $Mn^{3+}$ and $Mn^{4+}$, respectively [69, 70].

### 3.3 Morphological analysis

Figure S3 (a-b and c-d) (supplementary material) reveals the topographies of the as prepared ultrafine 1D α-MnO$_2$ nanorods through FESEM (Field emission electron microscopy) and TEM (Transmission electron microscopy), respectively. Highly uniform morphologies in 1D nano regime can be easily perceptible. Nanorods like morphologies can be assigned for the as prepared sample with diameters in the range of 10-40 nm.

Furthermore, in the process to obtain similar kind of ultrafine morphology for LiMn$_2$O$_4$, LiOH•H$_2$O were mixed with α-MnO$_2$ nanorods hydrothermally at 170 °C for 12h.

This resulted in a complete mixing of these two compounds. Figure 3 (a and b) shows FESEM images of the ultrafine 1D $α$-$MnO_2$ nanorods mixed with LiOH•$H_2O$. The low dimensionality of the nanorods allows to $α$-$MnO_2$ to be able to attract small nanoparticles of LiOH over their surfaces, which is due to their confined structure. It is well-known that nanoscale materials offer higher specific surface area than their bulk counterparts [54]. When these nanostructures are subject to hydrothermal pressure they easily get attracted towards other nanoparticles in order to reduce their surface area, and therefore, they are stabilized. Uniqueness arises, when the attracted particles also possess very low dimensional morphologies. This can well be visualized from the TEM images as shown in Figs. 3(c-d) within red-dotted circular portions. The as prepared unique mixture could have accelerated the solid state reaction during the process of achieving ultrafine 1D $LiMn_2O_4$ nanorods.

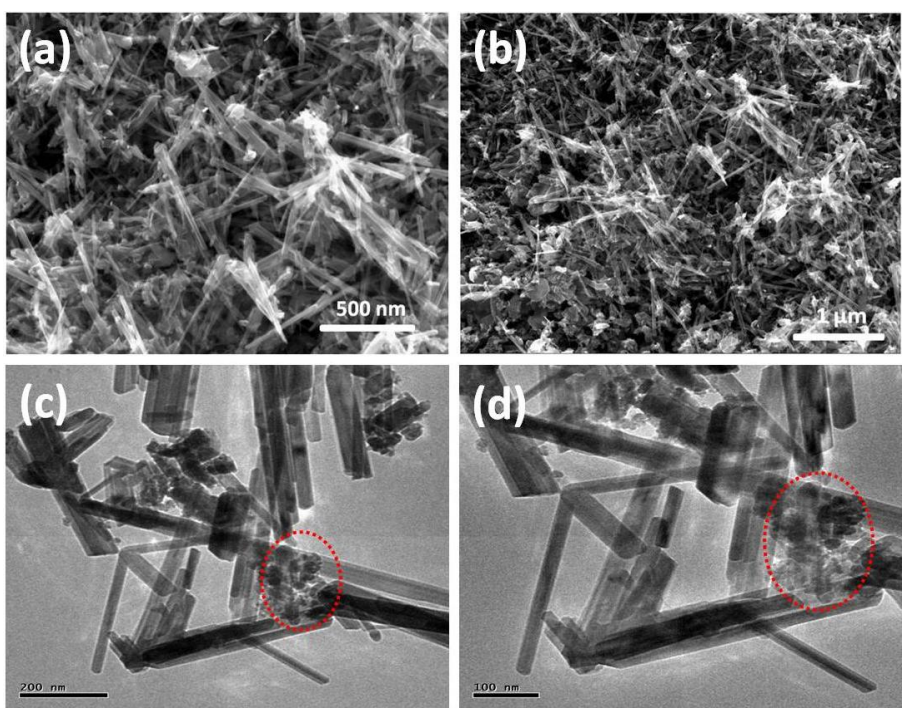

**Figure 3** (a-b) FESEM and (c-d) TEM images of the sample after hydrothermal treatment at 170 °C for 12 h using ultrafine 1D $α$-$MnO_2$ nanorods and LiOH•$H_2O$ as precursors.

This could be better visualized from the FESEM images of the sample prepared by solid state reaction at different temperatures of 100 and 400 °C shown in figure 4 (a-b and c-d), respectively. At 100 °C, the LiOH particles are supposed to settle over sharp edges of $MnO_2$ nanorods, making them stockier with perceptible physical deformations. The nanorods were quite successful in slimming down the morphologies of this added precursor (LiOH) by providing them with sharp surface edges for settlements. In this way they acted as template.

However, due to low temperature conditions, the solid state reaction between LiOH•H$_2$O and MnO$_2$ is supposed to get restricted, thereby resulting in stocky and interconnected rods like morphologies. When temperature was increased to 400 °C, the reaction got some meaning and rods with slimmer morphologies were viable. And finally, at 750 °C, ultrafine 1D LiMn$_2$O$_4$ nanorods were prevalent as shown in Fig. 5. A, though understanding of such growth process still requires further research.

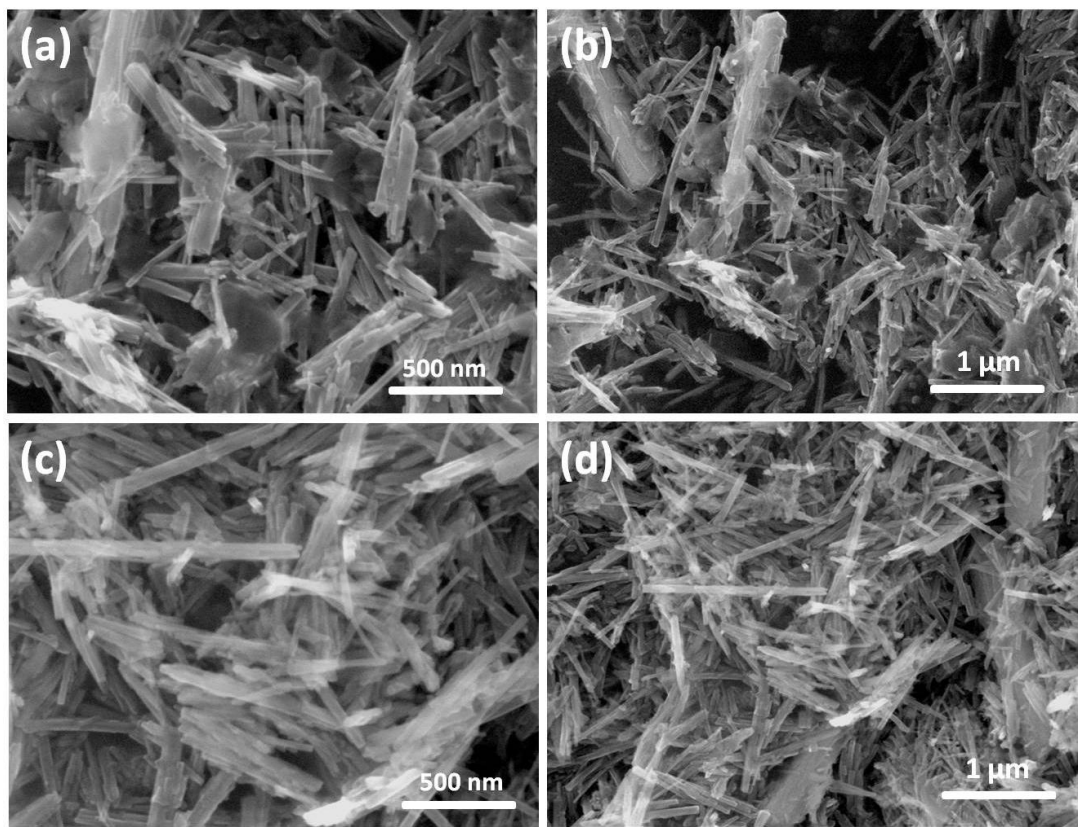

**Figure 4** FESEM images of sample prepared using ultrafine 1D *α*-MnO$_2$ nanorods and LiOH•H$_2$O as precursors at (a-b) 100 °C and (c-d) 400 °C.

Figure 5 (a-b and c-d) shows the FESEM and TEM images of the as prepared LiMn$_2$O$_4$ depicting conspicuous nanorods like morphology. The diameters of LiMn$_2$O$_4$ nanorods are about 10-50 nm. A wider view, presented in Fig. 5b, highlights that the as prepared sample is composed of only 1D morphologies since no other morphologies of higher dimensions is perceptible. It is important to note that both types of nanorods (*α*-MnO$_2$ and LiMn$_2$O$_4$) were free from any physical cracks and dislocations. Hence, ultrafine nature can be conferred for these nanorods. The nanorod morphology from the as obtained LiMn$_2$O$_4$ could be a deciding factor for enhancing its electrochemical performance as Li-ion battery cathode by offering short Li-ion diffusion paths. This achievement was only possible with the use of ultrafine 1D α-MnO$_2$ nanorods as precursor during the solid state reaction.

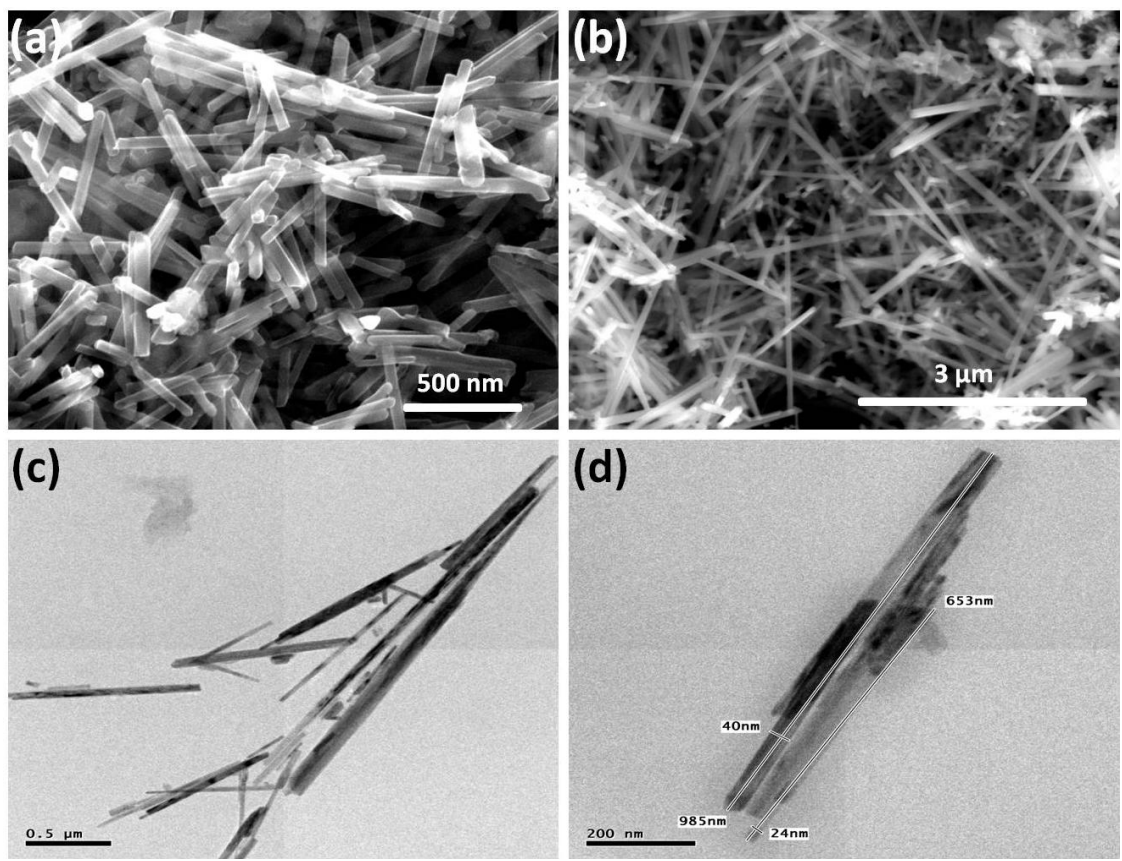

**Figure 5** (a-b) FESEM, (c-d) TEM images of the as prepared ultrafine 1D LiMn$_2$O$_4$ nanorods

HRTEM images, presented in Fig. 6 (a-d), show the ultrafine 1D LiMn$_2$O$_4$ nanorods wrapped inside the graphene oxide sheets. The nanorods seem to have diameters between 10-50 nm. These nanorods are clearly trapped inside the graphene oxide sheets, which act like a fishing net (red dotted border and circular area as shown in Fig. 6a). A promising morphology comprising an interconnected framework of graphene oxide nanosheet connecting the nanorods is seen clearly. In figs. 6(b-c), the dark shadowed portion resembles graphene oxide sheets enfolding the nanorods. It is expected that this type of morphology is favourable to increase the transport and interaction of charged particles at the interstitial sites of nanorods, which is highly desirable for enhanced electrochemical behaviour. Figure 6d displays a *d*-spacing of 0.48 nm, which is attributed to the indexed (111) plane of the cubic spinel structure of LiMn$_2$O$_4$ [71]. The visualization of uniform parallel lattice fringes suggests a high crystallinity for the as prepared ultrafine 1D LiMn$_2$O$_4$ nanorods. The observed interlayer *d*-spacing of 0.34 nm (Fig. 6d) accords well with the literature reported for the interlayer distance of graphene oxide sheets [72]. In corollary with the above structural and elemental analyses, a favourable surface interaction arising from Van der Walls interaction force can be predicted between the ultrafine 1D LiMn$_2$O$_4$ nanorods and the graphene oxide sheets. Such wrapping of the graphene oxide sheets can stabilize the volumetric change in the ultrafine 1D LiMn$_2$O$_4$ nanorods during lithiation as well as de-lithiation process, and decrease the Mn dissolution inside electrolyte, while concurrently improve the electron conductivity and reduce the Li-ion path diffusion, which are highly desirable for developing cathodic electrodes with high performance.

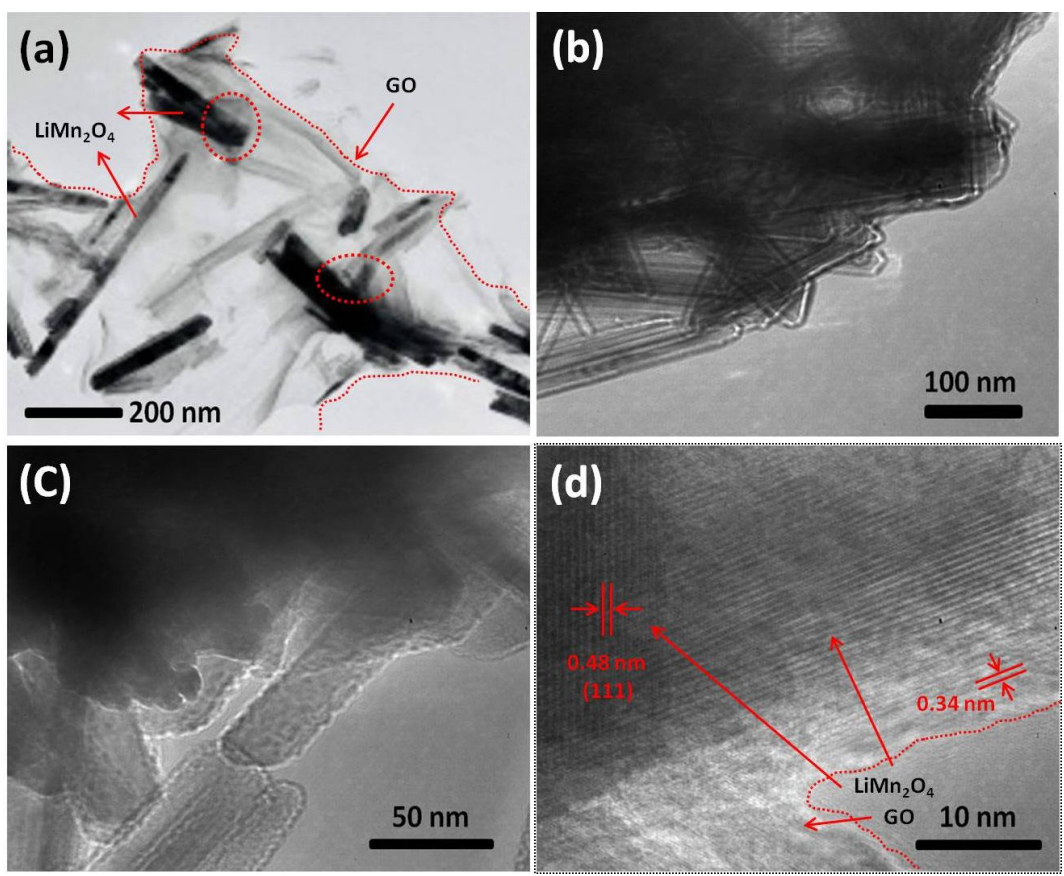

**Figure 6** (a-d) HRTEM images of LiMn$_2$O$_4$/GO nanocomposite with (d) *d*-spacing of 0.48 nm in (111) plane of the ultrafine 1D LiMn$_2$O$_4$ nanorods and 0.34 nm interlayer spacing of GO.

## 3.4 Electrochemical evaluation

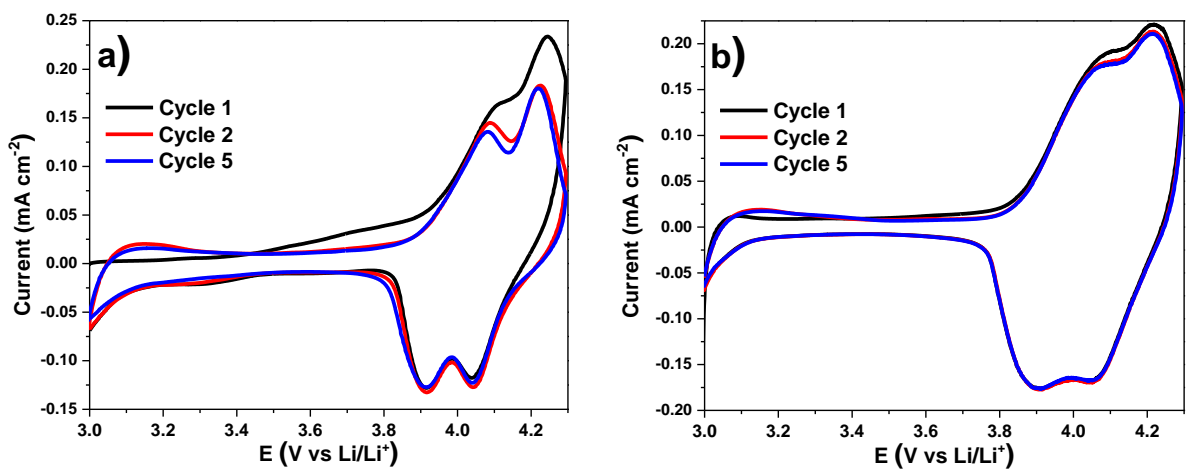

**Figure 7**. Cyclic voltammograms of: a) pristine ultrafine 1D LiMn$_2$O$_4$ nanorods and b) LiMn$_2$O$_4$/GO nanocomposite electrodes at 0.1 mV s$^{-1}$.

Figure 7 shows the cyclic voltammetry curves of both ultrafine 1D LiMn$_2$O$_4$ nanorods and LiMn$_2$O$_4$/GO nanocomposite electrodes in the potential window from 3.0 to 4.3 V vs Li/Li$^+$ at a scan rate of 0.1 mV s$^{-1}$ for first, second and fifth cycles. As shown in Fig. 7, during the first scanning cycle, there are two reversible redox peaks at 3.91/4.08 and 4.05/4.22 V vs Li/Li$^+$, which correspond to the two-step of insertion and de-insertion of Li$^+$-ions into and from the cubic spinel LiMn$_2$O$_4$ phase [73], which are indicative of the transformation between the cubic spinel phases (LiMn$_2$O$_4$ ↔ Li$_{1-x}$Mn$_2$O$_4$) and the coexistence of two phases (LiMn$_2$O$_4$ and Li$_{0.2}$MnO$_4$) [74], respectively. The irreversible broad peak between 3.6–3.9V vs Li/Li$^+$ in the first lithiation process is attributed to the solid electrolyte interphase (SEI) layer [75]. The well-overlapped peaks on the subsequent CV curves confirm an excellent cyclic performance of both ultrafine 1D LiMn$_2$O$_4$ nanorods and LiMn$_2$O$_4$/GO nanocomposite electrodes.

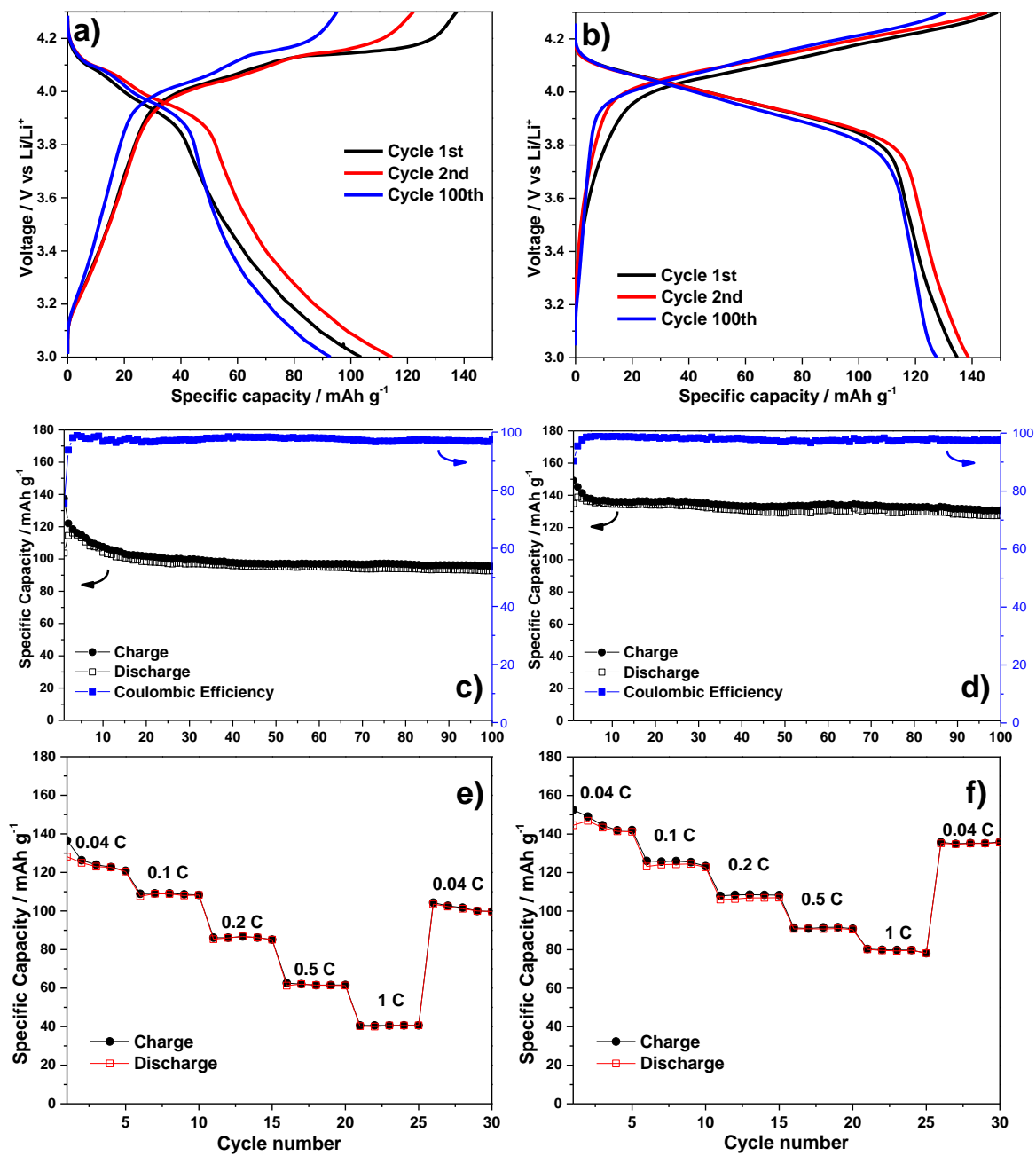

**Figure 8**. Electrochemical performance: (a, b) charge/discharge curves; (c, d) cycling performance at 0.05 C; (e, f) rate capability of the ultrafine 1D LiMn$_2$O$_4$ nanorod and LiMn$_2$O$_4$/GO nanocomposite electrodes, respectively.

To investigate the ultrafine 1D LiMn$_2$O$_4$ nanorods and LiMn$_2$O$_4$/GO nanocomposite electrochemical performance in terms of specific capacity, cyclability and rate capability, as lithium ion battery cathodes, these electrodes were cycled in a half-cell configuration. Figure (8a) and (8b) presents 1$^{st}$, 2$^{nd}$, 100$^{th}$ charge-discharge curves of the ultrafine 1D LiMn$_2$O$_4$ nanorods and LiMn$_2$O$_4$/GO nanocomposite electrodes between 3.0 and 4.3 V vs Li/Li$^+$ at a C-rate of 0.05 C, respectively. The charge and discharge profiles of both electrodes show the typical behavior associated to $LiMn_2O_4$ lithiation and de-lithiation processes [74, 76-78]. The initial specific charge capacity of the ultrafine 1D $LiMn_2O_4$ nanorods is 137 mAh g$^{-1}$ at C-rate 0.05 C, while the initial charge capacity is only 103 mAh g$^{-1}$, resulting in a relative low initial Coulombic efficiency of about 75%. However, when the graphene oxide sheets wrapped the ultrafine 1D $LiMn_2O_4$ nanorods, their electrochemical performance was improved, where the $LiMn_2O_4$/GO nanocomposite delivers the initial specific charge and discharge capacities of 149 and 134 mAh g$^{-1}$, respectively, with an initial Coulombic efficiency of about 90%. The high irreversible capacity loss and low Coulombic efficiency showed by both samples in the first cycle, it could mainly be the result for the conversion reaction, the electrolyte decomposition, and the formation of SEI. During the following cycles, the ultrafine 1D $LiMn_2O_4$ nanorods delivers a stable high specific charge capacity, 95 mAh g$^{-1}$ after 100 cycles with a Coulombic efficiency of ~97%, and initial charge capacity retention of ~70%. However, the $LiMn_2O_4$/GO nanocomposite shows a high specific charge capacity of 130 mAh g$^{-1}$ with a Coulombic efficiency of ~98% after 100 cycles, and initial charge capacity retention of ~87%. This implies that the graphene oxide layer improves the ultrafine 1D $LiMn_2O_4$ nanorod performance through an enhancement of the reversibility of the lithiation reaction during the (dis)charge processes.

The rate capability of the ultrafine 1D $LiMn_2O_4$ nanorods and $LiMn_2O_4$/GO nanocomposite were tested at various C-rates. As shown in Figure 8e, the ultrafine 1D $LiMn_2O_4$ nanorods electrode was charged and discharged at C-rate 0.04 C for 5 cycles at first, showing a fast capacity fading in the first few cycles to 126 mAh g$^{-1}$. Then, the C-rate was increased to 0.1 C for 5 cycles, and a reversible capacity of 109 mAh g$^{-1}$ was obtained. When the C-rate was changed to 0.25, 0.5 and 1 C, each one with intervals of 5 cycles, the corresponding reversible charge capacity was 86, 62 and 40 mAh g$^{-1}$, respectively. After changing the C-rate back to 0.04 C during 5 cycles intervals, the reversible capacity recovered to 101 mAh g$^{-1}$. However, the $LiMn_2O_4$/GO nanocomposite delivers reversible specific charge capacities of about 146, 125, 108, 91, 80 mAh g$^{-1}$, and recovers to 135 mAh g$^{-1}$ at C-rate of 0.04, 0.1, 0.25, 0.5, 1.0 C, and back to 0.04 C, respectively.

These results indicate that the excellent cyclic performance and high rate capability of the $LiMn_2O_4$/GO nanocomposite cathode could be associated with conductive wrapping of 2D graphene oxide sheets improving the Li$^+$-ion and electron diffusion into and from the ultrafine 1D $LiMn_2O_4$ nanorods. Moreover, it mitigated the agglomeration of these nanorods, when the lithiation and de-lithiation processes take place, allowing for maintaining a good electric contact between the ultrafine 1D LiMn$_2$O$_4$ nanorods in the electrode, and consequently resulting in a high capacity retention, stability and long-life performance.

## 4. Conclusions

Ultrafine 1D LiMn$_2$O$_4$ nanorods of 10-50 nm diameters were successfully grown with the help of as synthesized α-MnO$_2$ nanorods through a facile hydrothermal technique followed by a solid state reaction at 750°C. These nanorods, free from physical defects possessed highly pure cubic spinel LiMn$_2$O$_4$ phase. An enhanced electrochemical performance was achieved, when these 1D nanorods were effectively wrapped inside 2D graphene oxide sheets following the as developed facile technique of calcination at temperature of 750°C. The graphene oxide acts as an envelope to LiMn$_2$O$_4$ nanorods and established an interconnected framework thereby improving the 3D Li$^+$-ion transfer and electron diffusion. It also restricts the growth of the nanoparticles serving a conductive medium, thereby enhancing the cyclic stability and rate capability of the composite. The LiMn$_2$O$_4$/GO nanocomposite delivered a specific charge capacity of 130 mAh g$^{-1}$ at 0.05 C-rate after 100 cycles with a Coulombic efficiency of 98%, and initial specific charge capacity retention of 87%. Our analysis suggests that such excellent electrochemical performance is a result of conductive graphene oxide sheets wrapping on the crystalline ultrafine 1D LiMn$_2$O$_4$ nanorods surfaces, minimizing thereby Mn dissolution with favorable interfacial morphology.

## Acknowledgements

We acknowledge the DST Nano Mission, Govt. of India for the funding of the Project SR/NM/NS-1062/2012. We are also thankful for the support given by SENER-CONACYT, via Project No. 274314 to JR.## References